\documentclass{mem}
\usepackage{natbib}\usepackage{txfonts}\usepackage{balance}
\usepackage{graphicx}
\usepackage[a4paper]{hyperref}
\idline{75}{282}
\begin{document}
\def\teff{$T\rm_{eff }$}
\def\kms{$\mathrm {km s}^{-1}$}
\def\asca{{\it ASCA\/}}
\def\chandra{{\it Chandra\/}}
\def\conx{{\it Constellation-X\/}}
\def\heao1{{\it HEAO-1\/}}
\def\hst{{\it {\it HST}\/}}
\def\iras{{\it IRAS\/}}
\def\spitzer{{\it Spitzer\/}}
\def\scuba{{\it SCUBA\/}}
\def\rosat{{\it ROSAT\/}}
\def\rxte{{\it RXTE\/}}
\def\sax{{\it BeppoSAX\/}}
\def\xeus{{\it XEUS\/}}
\def\xmm{{XMM-{\it Newton\/}}}
\def\suzaku{{\it Suzaku\/}}
\def\swift{{\it Swift\/}}
\def\simbolx{{\it Simbol-X\/}}
\def\integral{{\it INTEGRAL\/}}

\title{The {\it SimbolX} view of the unresolved X--ray background}

   \subtitle{}

\author{
A. \,Comastri\inst{1},
R. \,Gilli\inst{1},
F. \,Fiore\inst{2}, 
C. \,Vignali\inst{3,1},
R. \,Della Ceca\inst{4}, 
\and G. \,Malaguti\inst{5}
          }

  \offprints{A. Comastri}

\institute{
INAF -- Osservatorio Astronomico di Bologna, Via Ranzani 1,
I-40127 Bologna, Italy  \email{andrea.comastri@oabo.inaf.it}
\and
INAF -- Osservatorio Astronomico di Roma, Via Frascati 33,
I-00040 Monteporzio Catone (RM), Italy
\and
Dipartimento di Astronomia, Universit\`a di Bologna, 
Via Ranzani 1, I-40127 Bologna, Italy
\and
INAF -- Osservatorio Astronomico di Brera, Via Brera 28,
I-20121 Milano, Italy
\and
INAF--IASF, via Gobetti 101,
I-40129 Bologna, Italy 
}

\authorrunning{Comastri et al.}

\titlerunning{SimbolX \& XRB}

\abstract{
We will briefly discuss the importance of sensitive X--ray observations 
above 10 keV for a better understanding 
of the physical mechanisms associated to the Supermassive Black Hole 
primary emission and 
to the cosmological evolution of the most obscured Active Galactic Nuclei.

}
\maketitle{}

\section{Introduction}

The fraction of the hard X-ray background (XRB) resolved into
discrete sources by deep {\it Chandra} and XMM-{\it Newton} observations 
smoothly decreases from practically  100\% below 2--3 keV 
to about 50\% in the 6--10 keV energy range (Worsley et al. 2005),
becoming negligible at energies above 10 keV, where the bulk of the 
energy density is produced.
\par
The shape and intensity of the unresolved XRB calculated 
by Worsley et al. (2005) is well described 
by a ``peaky" spectrum which is similar to that computed 
by folding the average spectrum of heavily obscured ($N_H > 10^{23}$ cm$^{-2}$) 
and Compton Thick ($N_H > 10^{24}$ cm$^{-2}$; hereinafter CT) AGN over the 
redshift range $z \sim$ 0.5--1.5 (e.g., Comastri 2004a). 
\par
The search for and the characterization of this population 
has become increasingly important in the last years and represents 
the last obstacle towards a complete census of accreting 
Supermassive Black Holes (SMBHs).  
According to the most recent version of XRB synthesis 
models (Gilli et al. 2007), heavily obscured and CT AGN are 
thought to be numerically as relevant as 
less obscured Compton Thin AGN. As a consequence, their impact 
on the astrophysics and evolution of AGN population as a whole 
cannot be neglected.
\par
It is convenient to consider absorption in the CT regime in two 
classes: mildly and heavily CT (see Comastri 2004b).
The column density of the absorbing gas of the former class 
is just above the
unity optical depth for Compton scattering, 
while it is much larger than unity for the latter. 
The primary radiation in mildly CT AGN is able to penetrate the 
obscuring gas and is visible above $\sim$ 10 keV, while Compton 
down-scattering mimics absorption over the 
entire energy range for heavily CT AGN. 
As far as the XRB is concerned, the most relevant contribution is 
expected to come from mildly CT AGN which are expected to be as bright 
as unobscured AGN above 10--15 keV. Though the flux of the primary emission 
is strongly depressed, also 
heavily CT AGN may provide some contribution 
to the XRB which depend on their scattering/reflection 
efficiency, most likely correlated with the geometry of the obscuring gas.
\par
At present, the best estimates of the basic properties of mildly 
CT AGN rely on the observations obtained with the {\tt PDS} 
instrument onboard {\it BeppoSAX} (Risaliti et al. 1999; 
Guainazzi et al. 2005). CT absorption, at 
least among nearby AGN, appears to be rather common. 
The relative fraction depends on the adopted selection 
criterium and can be as high as 50--60\% for optically selected 
Seyfert 2 galaxies.
More recently, {\tt BAT}/{\it Swift} (Ajello et al. 2007) 
and {\tt IBIS}/{\it INTEGRAL} (i.e., Bird et al. 2007) have surveyed 
the hard X--ray sky.
Although these surveys have provided relatively large numbers of 
hard X--ray sources (e.g., Krivonos et al. 2007), they are limited to 
bright fluxes ($S_{10-100 keV} > 10^{-11}$ erg cm$^{-2}$ s$^{-1}$),  
thus sampling only the very local Universe. As a consequence, the resolved
fraction of the hard ($>$ 10 keV) XRB is of the order of a few 
percent (Sazonov et al. 2007).
\par
Assuming the absorption distribution determined in the 
local Universe and folding it 
with cosmological evolution in AGN synthesis models, it is 
possible to predict the 
relative number of CT AGN in a purely hard X--ray selected sample.
Though the error bars suffer from small number statistics,  
there is a fairly 
good agreement with the model predictions (Figure~1), implying that CT 
absorption should be common also 
at high redshift and luminosities. However, only a few CT AGN 
are known beyond the local Universe (e.g. Comastri 2004b; 
Della Ceca et al. 2007).  
Even the deepest X--ray 
surveys in the 2--10 keV band (e.g., Tozzi et al. 2006;
Georgantopoulos et al. 2007) 
uncovered only a dozen of candidate heavily obscured CT AGN and await for 
better spectroscopic X--ray data for a confirmation. 
\par
It has been recently suggested that selection via mid--IR and 
optical colors is promising to pick up high-$z$ heavily obscured AGN 
(Fiore et al. 2007a; Daddi et al. 2007). Stacking the {\it Chandra} 
counts of sources selected on the basis of mid infrared (24$\mu$m) 
excess emission wrt the near--infrared optical flux, a strong 
signal is revealed in the hard band (up to 4--6 keV), which implies, 
at the average source redshift ($z$$\sim$2), $N_H$$\sim10^{24}$~cm$^{-2}$. 
Despite the similarities in the source selection and stacking 
techniques, the estimated space density of the candidate high-$z$ CT 
AGN differs by up a factor 3--5. The difference can be ascribed at least 
in part to the slightly different luminosity ranges sampled. 
According to Fiore et al. (2007a) the ``missing" population of luminous
($L_X > 10^{44}$ erg s$^{-1}$) candidate CT AGN at $z \sim$ 2 would 
have the right size to explain the 
{\it unresolved} XRB, while following Daddi et al. (2007) it may be 
significantly more numerous than predicted by synthesis models especially
at low ($L_X < 10^{43}$ erg s$^{-1}$) luminosities.  
The above estimates are affected by large uncertainties, however they 
strongly point towards the existence of a sizable population of accreting 
heavily obscured SMBHs at high redshift. 
The direct detection of the CT AGN population over the redshift range 
most densely populated by less obscured sources contributing 
to the XRB ($z \sim$ 0.7--1) bears important implications for  
our understanding of SMBH evolution. Moreover, a reliable determination 
of their luminosity function is a key parameter to reconcile the relic SMBH mass function 
with the local one estimated through the local 
$M_{BH} - \sigma / M_{BH} - M_{bulge}$ relationships and galaxies luminosity 
function (Marconi et al. 2004; Merloni et al. 2004). 

\begin{figure}[]
\resizebox{\hsize}{!}{\includegraphics[clip=true]{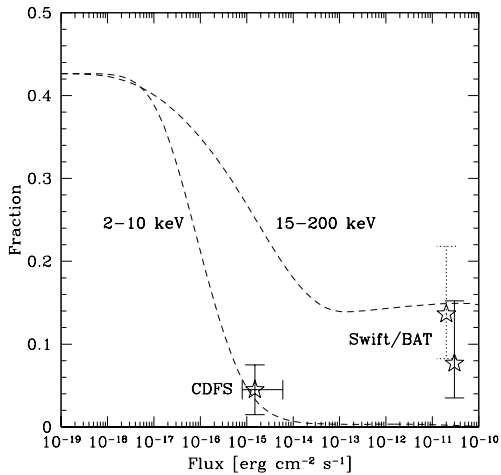}}
\caption{
\footnotesize The predicted fraction of CT AGN (adapted from Gilli et al. 2007) 
as a function of X-ray flux in two different energy ranges, as labeled. The 
points with error bars correspond to the estimates in the {\it Chandra} Deep Field 
South (CDFS; Tozzi et al. 2006) at faint fluxes,  
and from Swift/BAT survey (Markwardt et al. 2005). The upper point is computed
assuming that the unidentified sources are all CT AGN.
}
\label{pico}
\end{figure}

\section {Current hard X--ray observations}

The expected fraction of CT AGN in the 15--200 keV band as a function of flux 
is reported in Figure~1. The advantage of the hard X--ray selection wrt 
the 2--10 keV band is evident. 
The {\tt INTEGRAL} and {\tt Swift} all sky hard X--ray surveys have provided the first 
flux limited samples which can be compared with model predictions. 
There is a fairly good agreement, but the statistics is still dominated by fluctuations
associated to small numbers. Moreover, the CT nature of {\tt Swift} and {\tt INTEGRAL} 
sources
is estimated combining low signal to noise ratio hard X--ray ($>$ 10 keV) spectra 
with  observations at lower energies, mainly with {\it Chandra} and XMM--{\it Newton}. 
Follow--up observations of candidate CT AGN from the surveys described above  
were performed with {\it Suzaku}. The hard X--ray detector and the XIS CCD camera 
onboard the Japanese satellite are sensitive over a broad energy range 
($\sim$ 0.5--60 keV) and return good quality spectra for relatively bright 
hard X--ray sources.
Interestingly enough, {\it Suzaku} follow--up observations 
of hard X--ray selected {\tt INTEGRAL} and {\tt Swift} sources (Ueda et al. 2007; 
Comastri et al. 2007), 
only barely detected 
below 10 keV, suggest that a population of extremely hard sources, 
appearing only above 5--6~keV, may have escaped detection by surveys at lower energies.
Their high energy spectrum ($>$ 5--10 keV) is dominated by a strong reflection component 
from optically thick material, while there is no evidence for the presence of a
soft X--ray component, presumably due to scattering of the nuclear radiation, and common 
among Seyfert 2 galaxies. 
A possible explanation is that the geometrical distribution of the obscuring/reflecting  
gas is different from known Type~2 AGN.

By requiring that their integrated emissivity does not exceed the XRB level and associated 
uncertainties at 30 keV (e.g. Churazov et al. 2007;  Frontera et al. 2007), and assuming the 
same evolution of X--ray 
selected AGN (e.g. Hasinger et al. 2005; La Franca et al. 2005) it is, in principle, 
possible to constrain their number density. 
The hypothetic population of reflection dominated AGN could be up to a factor 4 more numerous 
than CT AGN (Figure ~2). 
Such an estimate is highly uncertain and strongly depends upon the assumption of a
reflection dominated spectrum.  
The bottom line of such an exercise is that there might be room for a 
previously unknown population of obscured AGN emerging only above 5--10~keV.

\begin{figure}[]
\resizebox{\hsize}{!}{\includegraphics[clip=true]{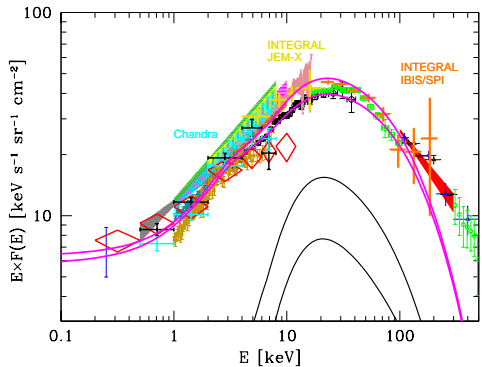}}
\caption{
\footnotesize The contribution of reflection dominated AGN in the Gilli et al. (2007) model  
(lower black line) 
can be increased up to a factor 4 (upper black line) and still be consistent with present 
uncertainties on the XRB level.
}
\label{frac}
\end{figure}

\section{Future imaging X--ray surveys} 
 
Imaging observations above 10 keV will offer a unique opportunity 
to address at least some of the issues mentioned above. The expected {\it SimbolX} 
capabilities are such to make possible the detection of a statistically 
significant sample of heavily obscured and CT AGN up to $z \sim$ 1. 
The expected quality of the spectral data over the range of fluxes accessible to 
{\it SimbolX} is extensively discussed in a companion paper (Della Ceca et al. 2007).

\begin{figure}[]
\resizebox{\hsize}{!}{\includegraphics[clip=true]{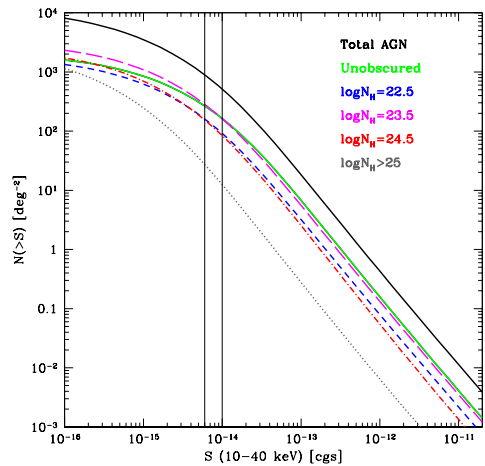}}
\caption{
\footnotesize The model predicted logN--logS in the 10--40 keV band. The number counts 
for different absorption column densities are reported with the labeled colour code.
The two vertical lines correspond to the expected limiting fluxes for the present
SimbolX configuration: $\sim 6 \times 10^{-15}$ cgs on-axis and $\sim$ 10$^{-14}$ cgs 
for an off-axis angle of 5 arcmin.
}
\label{frac}
\end{figure}

The expected number counts in the 10--40 keV band (Figure~3) are obtained extrapolating 
the spectral energy and absorption distributions of the Gilli et al (2007) 
synthesis model.
The hard X--ray source surface density keeps a steep (close to Euclidean) slope
down to the expected sensitivity limits of {\it SimbolX} observations 
($\sim 10^{-14}$ erg cm$^{-2}$ s$^{-1}$) 
in the 10--40 keV band.  
More specifically, a few hundreds CT  AGN
per square degree are predicted at the limits of a deep survey.
By scaling the estimated space densities for the {\it SimbolX} 
field of view, our estimates 
translate in a few up to about ten CT AGN in a deep (1 Msec) pointing. 
Taking into account the off-axis sensitivity and the fact that the faintest 
sources will be detected with a number of counts insufficient for X--ray spectral analysis, 
the search for the most obscured AGN cannot be pursued only with 
deep pointings. 
A comprehensive understanding of the physics and evolution of distant and 
obscured AGN, as well as of different classes of cosmic sources, requires
a close synergy between deep and wide area surveys. 
Indeed, the evolution
of X--ray selected AGN was determined by a large number of surveys which 
have extensively sampled the solid angle vs. depth discovery space 
(see Fig.~1 of Brandt \& Hasinger 2005). 
A possible strategy is discussed in the Fiore et al (2007b) companion paper.
\par
The predicted redshift distribution of CT AGN at three different limiting 
fluxes is reported in Figure~4. Not surprisingly, the peak moves to higher redshifts 
as the sensitivity increases. 
A pronounced  high redshift ($z \sim$ 0.5--2) tail 
starts to develop at fluxes around 10$^{-13}$ erg cm$^{-2}$ s$^{-1}$ and lower.
As far as the search for obscured accretion at high redshift is concerned,  
the best trade--off would be to maximize the area covered close
these fluxes.

\begin{figure}[]
\resizebox{\hsize}{!}{\includegraphics[clip=true]{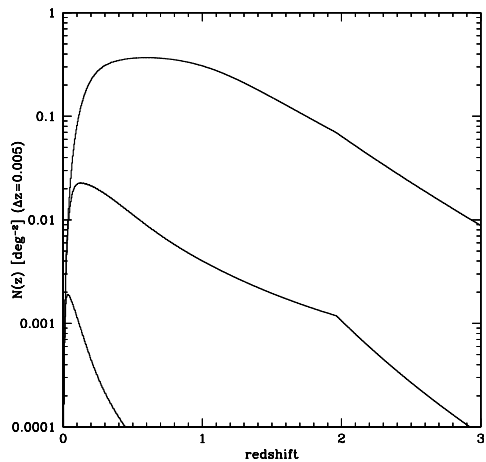}}
\caption{
\footnotesize The redshift distribution of CT AGN for three different limiting fluxes 
($10^{-12} - 10^{-13} - 10^{-14}$ erg cm$^{-2}$ s$^{-1}$) from bottom to top. 
}
\label{frac}
\end{figure}

\section{Conclusions}

Deep {\it SimbolX} surveys will be able to resolve about half of the XRB in the 10--40 
keV energy range (Figure~5). This achievement will represents a major step forward.
In fact,  at present, less than a few percent of the XRB in that band is resolved.
A comparison with an extrapolation of an AGN synthesis model (Comastri 2004a) 
tuned to reproduce the resolved background below 10 keV is reported in Fig~5. 
Although the level of the XRB predicted to be resolved by {\tt SimboX} 
is close to the model extrapolation, it will be possible, with the present configuration, 
to probe the entire range of absorption of the XRB sources.
Several ``new"  CT AGN (see Fiore et al. 2007b), which are undetected 
even in the deepest  XMM--{\it Newton} and {\it Chandra} surveys (Fig.~1), will be revealed.
Moreover, given that the fraction of CT AGN is expected to steeply increase just below 
$\sim 10^{-14}$ erg cm$^{-2}$ s$^{-1}$, it would be highly rewarding 
to push the limiting flux to somewhat lower fluxes.
It is worth remarking that the resolution of a significant fraction 
of the XRB  has always brought significant advances in our understanding of AGN evolution.
{\it ROSAT} deep surveys showed that luminous unobscured quasars at high redshift ($z \sim$ 1.5--2) 
were responsible for most of the soft (around 1 keV) XRB (Lehmann et al. 2001). 
Later on, thanks to {\it Chandra} and XMM--{\it Newton}
deep surveys, it was demonstrated that the bulk of the XRB at least up to 5--6 keV is originating
in relatively low luminosity sources, most of them obscured, at $z \sim$ 1 (Brandt \& Hasinger 2005).
While we expect to uncover the so far elusive population of CT AGN  up to relatively 
high redshift, it may well be possible that the content of the X--ray sky 
above $\sim$ 10 keV is different from what predicted. 
We look forward to new unexpected findings which could be obtained by pushing imaging 
observations in the so far unexplored hard X--ray band.

\begin{figure*}[t!]
\resizebox{0.98\hsize}{!}{\includegraphics[clip=true]{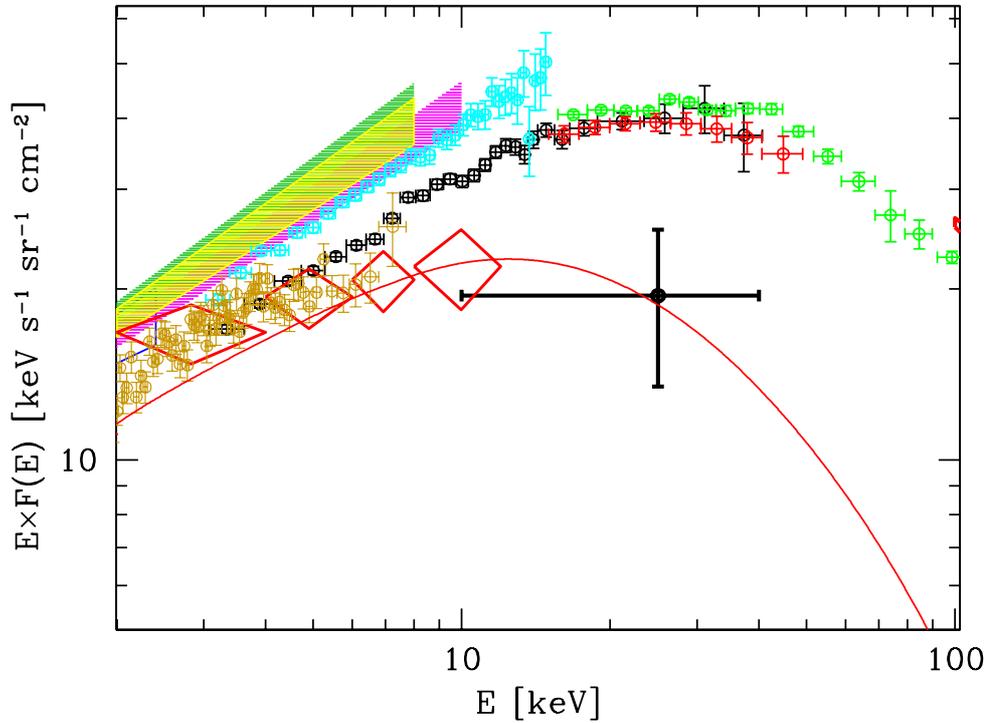}}
\caption{\footnotesize
The fraction of the 10--40 keV XRB which will be resolved by a deep 1 Ms {\it SimbolX} pointing
(point with error bars). The XRB measurements  below 10 keV are from {\it Chandra} and 
XMM--{\it Newton} 
as well as the resolved fraction in deep fields (gold points and red squares).
while the points extending up to $\sim$ 15 keV are from RXTE (Revnivtsev et al. 2003). 
At high energy, the HEAO1--A4 spectrum (Gruber et al. 1999; green points) is reported along 
with the recent \sax\ PDS observations (Frontera et al. 2007; red points).
The red solid line represents an AGN synthesis model (see Comastri 2004a) which reproduces
the observed level of the resolved XRB below 10 keV. 
}
\label{eta}
\end{figure*}

\begin{acknowledgements}
We acknowledge financial contribution from contracts ASI--INAF I/023/05/0, ASI--INAF I/088/06/0 
and PRIN--MUR grant 2006--02--5203.
\end{acknowledgements}

\bibliographystyle{aa}

\end{document}